\documentclass{aa}

\usepackage{graphicx}
\usepackage{txfonts}
\usepackage{textcomp}
\usepackage{natbib}
\bibpunct{(}{)}{;}{a}{}{,}

\usepackage{color}

\begin{document}

   \title{Broad-band spectrophotometry of the hot Jupiter HAT-P-12b from the near-UV to the near-IR}
   \titlerunning{Broad-band spectrophotometry of HAT-P-12}


   \author{M. Mallonn \inst{1}, V. Nascimbeni \inst{2}, J. Weingrill \inst{1}, C. von Essen \inst{3}, K. G. Strassmeier \inst{1}, G. Piotto \inst{2,4}, I. Pagano \inst{5}, G. Scandariato \inst{6}, Sz. Csizmadia \inst{7}, E. Herrero \inst{8}, P. V. Sada \inst{9}, V. S. Dhillon \inst{10}, T. R. Marsh \inst{11}, A. K\"{u}nstler \inst{1}, I. Bernt \inst{1}, T. Granzer \inst{1}}

   \authorrunning{M. Mallonn et al.}

   \institute{Leibniz-Institut f\"{u}r Astrophysik Potsdam, An der Sternwarte 16, D-14482 Potsdam, Germany\\
              \email{mmallonn@aip.de}
         \and
             INAF - Osservatorio Astronomico di Padova, Vicolo dell’Osservatorio 5, Padova, IT-35122, Italy 
         \and
             Stellar Astrophysics Centre (SAC), Department of Physics and Astronomy, Aarhus University, Ny Munkegade 120, DK-8000 Aarhus C, Denmark
          \and
             Dipartimento di Fisica e Astronomia “Galileo Galilei”, Universit`a di Padova, Vicolo dell’Osservatorio 3, Padova IT-35122, Italy
         \and
             INAF – Osservatorio Astrofisico di Catania, via S. Sofia 78, 95123 Catania, Italy
         \and
             INAF – Osservatorio Astronomico di Palermo, Piazza del Parlamento 1, 90134 Palermo, Italy
         \and
             Institute of Planetary Research, German Aerospace Center, Rutherfordstrasse 2, D-12489 Berlin, Germany
         \and
             Institut de Ciències de l’Espai (CSIC–IEEC), Campus UAB, Facultat de Ciències, Torre C5 parell, 2a pl, 08193 Bellaterra, Spain
         \and
             Universidad de Monterrey, Departamento de Física y Matemáticas, Avenida I. Morones Prieto 4500 Poniente, San Pedro Garza García, Nuevo León, 66238, México
         \and
             Department of Physics and Astronomy, University of Sheffield, Sheffield S3 7RH, UK
         \and
             Department of Physics, University of Warwick, Coventry CV4 7AL, UK\\
             }

   \date{Received ; accepted }

 
  \abstract
   {The detection of trends or gradients in the transmission spectrum of extrasolar planets is possible with observations at very low spectral resolution. Transit measurements of sufficient accuracy using selected broad-band filters allow for an initial characterization of the atmosphere of the planet.}
   {We want to investigate the atmosphere of the hot Jupiter HAT-P-12b for an increased absorption at the very blue wavelength regions caused by scattering. Furthermore, we aim for a refinement of the transit parameters and the orbital ephemeris. }
   {We obtained time series photometry of 20 transit events and analyzed them homogeneously, along with eight light curves obtained from the literature. In total, the light curves span a range from 0.35 to 1.25 microns. During two observing seasons over four months each, we monitored the host star to constrain the potential influence of starspots on the derived transit parameters.}
   {We rule out the presence of a Rayleigh slope extending over the entire optical wavelength range, a flat spectrum is favored for HAT-P-12b with respect to a cloud-free atmosphere model spectrum. A potential cause of such gray absorption is the presence of a cloud layer at the probed latitudes. Furthermore, in this work we refine the transit parameters, the ephemeris and perform a TTV analysis in which we found no indication for an unseen companion. The host star showed a mild non-periodic variability of up to 1\%. However, no stellar rotation period could be detected to high confidence.}
   {}


   \maketitle
%

\section{Introduction}

Transit measurements at multiple wavelengths allow for measuring the so-called transmission spectrum of an extrasolar planet. It consists of the search for a wavelength dependence of the apparent radius of an exoplanet. At wavelengths of high atmospheric opacity, the atmosphere becomes optically thick at higher altitudes for the transmitted star light than at wavelengths of lower opacity. Such measurements are not limited to spectroscopic transit observation. In fact, spectrophotometry of transits can be done using multiple broad-band filters to create a ``spectrum'' that consists of a few data points over the optical and near infrared (NIR) wavelength range. Similar to the color index of stars, which allows for a rough classification of stars, it might be possible to distinguish between several categories of planetary atmospheres by using only a handful well-separated color filters \citep{Nascimbeni2013,Bento2014}.

The required accuracy when measuring the relative differences in the apparent planetary radius can be achieved on small telescopes by using broad-band filters. Str\"{o}mgren or even narrower filter systems in combination with a small telescope aperture would yield too high a photon noise in the light curves. However, the use of broad filters like the Johnson/Bessel or Sloan filter system comes with the drawback that signals of narrow spectral features in the planetary atmosphere are strongly damped and smeared out. For example, detecting absorption from the sodium D line becomes more difficult. Instead, broad-band observations can be used to search for general trends in the spectrum, such as the rising absorption toward short wavelengths caused by Rayleigh-scattering or the predicted redward slope between very blue colors and the central part of the optical window caused by TiO \citep{Fortney2010}. Despite the growing number of observational publications in very recent years, the task remains 
challenging because of the weakness of the expected signals.

The atmospheric properties of several exoplanets have been constrained with ground-based broad-band observations. \cite{Nascimbeni2013} and \cite{Biddle2014} show that the spectrum of the warm Neptune GJ3470b cannot be explained with a gray, featureless atmosphere. Instead, a Rayleigh slope needs to be included in the modeling to gain a reasonable fit to the observed data. For the atmosphere of the super-Earth GJ1214b, \cite{deMooij2012} ruled out a cloud-free solar composition, and \cite{Nascimbeni2015} provided evidence against Rayleigh scattering at blue wavelengths. \cite{Southworth2014b} found a strong increase in absorption toward the blue for the hot Jupiter WASP-103b, which is currently unexplained. The achieved precision of the broad-band spectrophotometry papers of \cite{Bento2014} and \cite{Mancini2013c} did not yield significant results for the hot Jupiters WASP-17b and WASP-19b, but tentatively it was possible to favor a certain atmospheric model before the others. In contrast, there is a larger number of broad-band spectrophotometry investigations of hot Jupiters that did not reach the necessary precision to distinguish between atmospheric models \citep{Southworth2012,Nikolov2013,Mancini2013b,Mancini2014,Mancini2014b,Copperwheat2013,Chen2014,Fukui2014,Zhou2014}.

In this work, we investigate the hot Jupiter HAT-P-12b. It was discovered in 2009 by \cite{Hartman2009} and has a mass of $\sim$\,0.2~M$_J$ and a radius of about 1~R$_J$. The host star HAT-P-12 is a K5 main sequence star with a mass of 0.73~M$_\odot$, a radius of 0.7~R$_\odot$, and a magnitude of $\mathrm{V}\,=\,12.8$~mag \citep{Hartman2009,Ehrenreich2011}. The planetary mean density and the surface gravitational acceleration is rather low ($g \sim 6$~ms$^{-2}$). This low surface gravity and the equilibrium temperature of $\sim$\,1000~K result in an atmospheric pressure scale height of about 600~km. Together with the large transit depth due to the small stellar size, HAT-P-12b is among the known exoplanets with the strongest potential transmission signal. The planet orbits its host star every $\sim$\,3.2~days. HAT-P-12 was described as moderately active \citep{Knutson2010}. However, the jitter of the RV measurements is low with a few meters per second \citep{Hartman2009}. The line cores of the Ca H\&K lines 
show emission features with log($R'_{HK}$)\,=\,--5.1. 

Follow-up photometry of HAT-P-12b was published by \cite{Sada2012}, who present one J band transit light curve. Shortly after, three optical light curves were analyzed by \cite{Lee2012}, allowing for a transit parameter refinement and a more accurate ephemeris. \cite{Todorov2013} published upper limits on the eclipse depth and the brightness temperature at 3.6 and 4.5~$\mu$m. A first transmission spectrum was presented by \cite{Line2013} using HST in the NIR.

The large number of transit light curves analyzed in this work provide the opportunity to analyze the data for transit timing variation (TTV). This technique allows the detection of so far unseen planetary mass companions and derivation of various system parameters \citep{Agol2005}. In general, the current understanding of planet formation and migration predicts that hot Jupiters orbit their host stars alone \citep[see,  e.g.,][and references therein]{Steffen2012}. However, there might be exceptions from this general rule, as when \cite{Szabo2013} presented several exoplanet candidates detected by TTVs of hot Jupiters based on Kepler data. Furthermore, \cite{Maciejewski2013} and \cite{vonEssen2013} have announced tentative detections of TTV signals using ground-based data. However, ground-based results should be used with care as the debate for TTVs in the WASP-10 system illustrates \citep{Maciejewski2011,Barros2013}. There are also a handful of examples of candidates or confirmed multiplanet systems with a hot Jupiter, where indications or evidence of another companion was achieved with the radial velocity method, see \cite{Hartman2014} and references therein. \cite{Lee2012} found variations in the transit times of HAT-P-12b that are significantly larger than their error values. We use our sample of transit observations to follow up on this potential TTV detection.

Transit light curves always suffer to some degree from the correlated noise caused by the Earth's atmosphere or instrumental effects \citep{Pont2006}. One approach to mitigating this is to observe transits at multiple epochs and multiple telescopes since these light curves are not correlated to each other, and the influence of the individual red noise averages out \citep{Lendl2013}. In this work we combine a large number of transits in their majority taken with small-to-medium class telescopes at different epochs to refine the ephemeris and orbital parameters and to search for transit timing variations and a wavelength dependence of the planetary radius. 
Sections~2 and 3 describe the observations and the data reduction. Section~4 presents the results of the re-analysis of the literature data. The analysis and results of the new transit light curves are given in Section~5, while Section~6 details the results of the monitoring program. Section~7 discusses the results, followed by the conclusions in Section~8.

\section{Observations}
We acquired data of 20 primary transit events. Observations were done with the STELLA telescope, the Large Binocular Telescope (LBT), the William Herschel Telescope (WHT), the Telescopio Nazionale Galileo (TNG), the Nordic Optical Telescope (NOT), the Asiago 1.82m telescope, and the 0.8m telescope of the Montsec Astronomical Observatory (OAdM). The transits of the LBT and the WHT were observed in multichannel mode. The STELLA telescope has been used to monitor the host star HAT-P-12 in two colors for a length of about four months in 2012 and four months in 2014. A summary of the transit observations is given in Table \ref{tab_overview} including the date of observation, filter, exposure time, time between consecutive exposures (cadence), number of data points, the rms, the $\beta$ factor (see Section \ref{chap_anal}), and the airmass range during the observation.

\subsection{Transit observations}

\subsubsection{STELLA}
STELLA observed 13 transits in the observing seasons 2011, 2012, and 2014. STELLA is a robotic 1.2m twin telescope located on Tenerife \citep{Strassmeier2004}. The instrument of use was its wide-field-imager WiFSIP \citep{Weber2012} with a field of view (FoV) of 22'$\times$22' on a scale of 0.32"/pixel. The detector is a single 4096$\times$4096 back-illuminated thinned CCD with 15$\mu$m pixels. The telescope was slightly defocused during the observations (FWHM typically 2-3 arcseconds) to minimize flat-fielding errors (see, e.g., \citealt{Southworth2009}). The observations have been carried out using the Sloan r' filter in the 2011 season and quasi-simultaneously with alternating filters Johnson B and V in 2012. The B band light curves are of low quality because of the intrinsic faintness of HAT-P-12 at blue wavelengths and weather problems, therefore they are not analyzed here. In the year 2014, two transits were observed in Sloan g' and one in Sloan i'. The r' observations from 2011 were performed reading 
out the entire detector area. In the season 2012, we optimized the setup with the read-out of a central window of 1k by 1k to save read-out time. Typically, the object centroid drifted by a few pixels per hour over the detector.

\subsubsection{LBT}
One transit was observed with the Large Binocular Camera \citep[LBC, ][]{Giallongo2008} on April 4, 2013 (PI: I. Pagano). The LBC consists of two prime-focus, wide-field imagers mounted on the left and right arms of the LBT, and is optimized for blue and red optical wavelengths, respectively. We applied the U$_{spec}$ filter on the blue arm, a filter with Sloan u' response but having an increased efficiency, centered at $\lambda_c = 357.5$~nm, and the F972N20 filter on the red arm, which is an intermediate-band filter centered at $\lambda_c = 963.5$~nm. The two filters have been chosen to cover a very wide wavelength range; furthermore, the F972N20 filter avoids most of the telluric lines in the red region. On both sides, just the central chip of four chips has been read out to save hard disk space and a region of interest (window) has been defined to save read-out time. The final FoV encompassed $\sim$\,7.5$\times$7.5~arcminutes$^2$. The blue arm was defocused to an artificial FWHM of about four~arcseconds, 
the PSF 
of the red side was donut-shaped with a width of about eight~arcseconds at half the peak count rate. The exposure time was set to 90~seconds on both sides with a cadence of $\sim$\,125~seconds on the blue and $\sim$\,117~seconds on the red side. Unfortunately, the observation was affected by clouds, especially in the second half of the transit, almost no flux of the target was received. However, the observation was not interrupted, and a number of useful frames were again recorded after egress. The telescope was passively tracking during the time series, resulting in a centroid drift of about 20~pixels in both directions on the blue side and 80/40~pixels in x/y on the red side.

\subsubsection{WHT}
We observed one transit on March 15, 2014 (PI: V. Nascimbeni) with the triple beam, frame-transfer CCD camera ULTRACAM \citep{Dhillon2007}. The instrument splits the incoming light into three channels with individual CCDs, allowing simultaneous observations with three filters. We chose the Sloan filters u', g', and r'. The exposure time for all three channels was 5.7~seconds with almost no overheads (25~ms) owing to the frame-transfer technique. The telescope has been defocused to a FWHM of approximately 5/4/3.5~arcseconds in the u'/g'/r' band. The observing conditions were photometric. The object centroids moved on the CCD over the time series by about five~pixels in u' and less than two pixels in g' and r'. ULTRACAM shows no measurable flexure, therefore we attribute the centroid drift to atmospheric refraction. No windowing was applied to the FoV of 5$\times$5~arcminutes$^2$. For the u' channel, four exposures were co-added on chip before read-out to reduce the influence of the read-out noise.

\subsubsection{TNG}
One transit was observed with the TNG on May 28, 2014 using the focal reducer DOLORES in imaging mode (DDT program, PI: M. Mallonn). DOLORES is equipped with a 2048$\times$2048 E2V 4240 CCD camera optimized for blue wavelengths and a FoV of 8.6$\times$8.6~arcminutes$^2$. We employed the permanently mounted Sloan u' filter. A binning of 2$\times$2 pixels was used to reduce the read-out time. An exposure time of 40~seconds resulted in a cadence of 49~seconds. A mild defocus was applied throughout the time series to reach an artificial FWHM of about two~arcseconds. The observing conditions were non-photometric with variable seeing and the target flux varied by up to 30\% on a short time scale. Twice during the time series, the position of the target jumped abruptly by about ten~pixels. In between these jumps the object drifted by less than three~pixels/hour.

\subsubsection{NOT}
One transit was observed with the NOT on March 15, 2014 as a Fast-Track program (PI: M. Mallonn). The imaging time series was taken with the stand-by CCD imager StanCam, which hosts a 1k$\times$1k TK1024A CCD detector. The FoV amounted to 3$\times$3~arcminutes$^2$, sufficient to image the target and one nearby reference star of similar brightness. We used a Bessel B filter with an exposure time of 60~seconds, resulting in a cadence of 103~seconds. The night was photometric with a seeing between 1.0 and 1.3~arcseconds. We noticed an object drift of about six~pixels in both directions over the length of the time series.

\subsubsection{Asiago}
Two transits were observed with the Asiago 1.82m telescope (PI: G. Piotto) and its Asiago Faint Object Spectrograph and Camera (AFOSC). The camera hosts a 2k$\times$2k E2V 42-20 CCD, which images the 9$\times$9~arcminutes$^2$ FoV with a 0.26''/pixel plate scale. We applied a 4$\times$4~pixel binning and read out only a region of interest to reduce the overheads to about two~seconds. All images were exposed for ten~seconds. A defocus was used, which enlarged the PSF to a three to five~arcsecond FWHM.

\subsubsection{OAdM}
One partial transit was observed on May 15, 2014 (PI: E. Herrero) with the fully robotic 80-cm Ritchey-Chrétien telescope of the Montsec Astronomical Observatory \citep{Colome2008}. The telescope hosts a FLI PL4240 2k$\times$2k camera with a plate scale of 0.36 arcseconds per pixel. Because of the intrinsic faintness of the K dwarf HAT-P-12 at blue wavelengths, we chose the reddest filter available here, the Cousin I filter. The exposure time amounted to 120~seconds with a cadence of 136~seconds. The beginning of the astronomical twilight prevented us from observing a full transit. The telescope was mildly defocused to an FWHM of $\sim$\,2.5~arcseconds, the centroids drifted over the time series by 60 and 30~pixels in the x and y directions.

\begin{table*}
\caption{Overview about new transit observations of HAT-P-12b. The columns give the observing date, the used telescope, the chosen filter, the exposure time, the observing cadence, the number of observed individual  data points, the dispersion of the data points as root-mean-square (rms) of the observations after subtracting a transit model and a detrending function, the $\beta$ factor (see Section \ref{chap_anal}), and the airmass range of the observation.}
\label{tab_overview}
\begin{center}
\begin{tabular}{p{20mm}p{20mm}p{10mm}p{10mm}p{10mm}p{10mm}p{10mm}p{10mm}l}
\hline
\hline
\noalign{\smallskip}


Date        & Telescope &  Filter &  $t_{\mathrm{exp}}$ (s) & Cadence (s) & $N_{\mathrm{data}}$ &  rms (mmag) &  $\beta$ &  Airmass \\
\hline
\noalign{\smallskip}
2011 Mar 9   &  Asiago   &  R             &   10   &   12.1   &   1117  &  2.00    &   1.12  &    1.00-1.20  \\
2011 Apr 7   &  STELLA   &  r'            &   40   &   78     &    205  &  1.92    &   1.64  &    1.03-1.91  \\
2011 Apr 23  &  STELLA   &  r'            &   40   &   78     &     85  &  1.45    &   1.00  &    1.07-1.28  \\
2011 May 9   &  STELLA   &  r'            &   40   &   78     &    204  &  1.26    &   1.02  &    1.03-1.59  \\
2011 May 25  &  STELLA   &  r'            &   40   &   78     &    204  &  1.83    &   1.00  &    1.08-2.77  \\
2011 Jun 7   &  STELLA   &  r'            &   40   &   78     &    205  &  1.14    &   1.00  &    1.03-1.61  \\
2011 Jun 23  &  STELLA   &  r'            &   40   &   78     &    203  &  2.03    &   1.06  &    1.07-2.76  \\
2012 Feb 6   &  STELLA   &  V             &   30   &   145    &    133  &  4.10    &   1.08  &    1.03-1.90  \\
2012 Mar 19  &  Asiago   &  R             &   10   &   12.1   &    930  &  2.11    &   1.11  &    1.03-1.42  \\
2012 May 3   &  STELLA   &  V             &   40   &   186    &     92  &  2.04    &   1.03  &    1.03-1.36  \\
2012 May 6   &  STELLA   &  V             &   40   &   186    &    107  &  2.77    &   1.05  &    1.03-2.02  \\
2012 June 4  &  STELLA   &  V             &   40   &   128    &    129  &  3.88    &   1.00  &    1.03-2.12  \\
2013 Apr 5   &  LBT      &  U$_{spec}$    &   90   &   125    &     66  &  1.57    &   1.18  &    1.02-1.28  \\
2013 Apr 5   &  LBT      &  F972N20       &   90   &   117    &     63  &  0.98    &   1.00  &    1.02-1.28  \\
2014 Feb 27  &  STELLA   &  g'            &   80   &   104    &    140  &  1.69    &   1.00  &    1.07-2.17  \\
2014 Mar 15  &  STELLA   &  g'            &   80   &   104    &    136  &  1.34    &   1.45  &    1.03-1.28  \\
2014 Mar 15  &  NOT      &  B             &   60   &   103    &    139  &  1.52    &   1.00  &    1.04-1.34  \\
2014 Mar 15  &  WHT      &  u'            &   23   &   23     &    676  &  7.16    &   1.36  &    1.03-1.38  \\
2014 Mar 15  &  WHT      &  g'            &   5.7  &   5.7    &   2717  &  1.69    &   1.96  &    1.03-1.38  \\
2014 Mar 15  &  WHT      &  r'            &   5.7  &   5.7    &   2717  &  1.27    &   2.00  &    1.03-1.38  \\
2014 May 15  &  OAdM     &  I             &   120  &   136    &     40  &  1.44    &   1.76  &    1.20-1.51  \\
2014 May 28  &  STELLA   &  i'            &   45   &   72     &    240  &  1.55    &   1.00  &    1.04-1.56  \\
2014 May 28  &  TNG      &  u'            &   40   &   49     &    308  &  5.44    &   1.22  &    1.03-1.32  \\
\hline                                                                                                     
\end{tabular}
\end{center}
\end{table*}

   \begin{figure*}
   \centering
   \includegraphics[width=16cm]{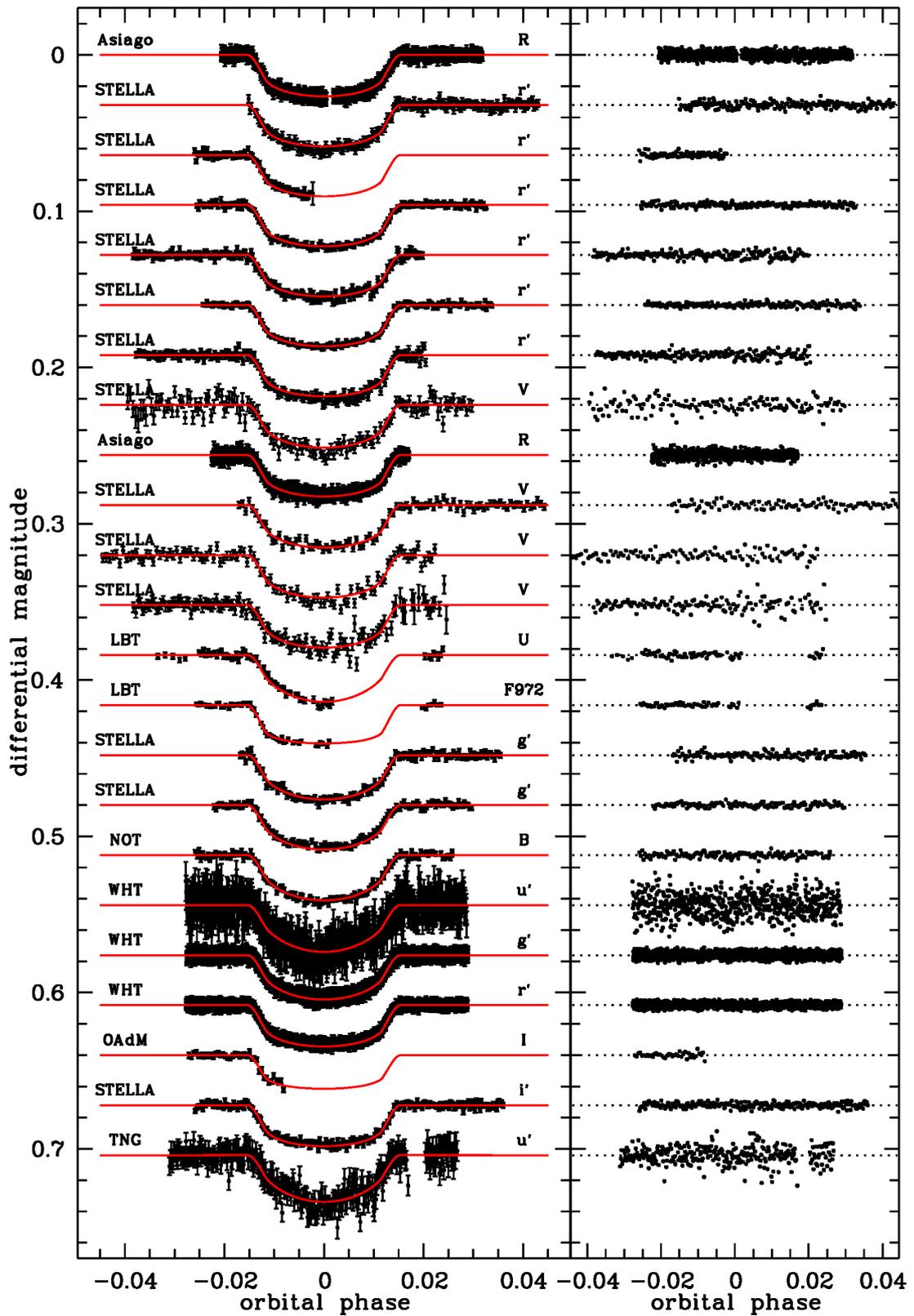}
       \caption{Light curves of Table \ref{tab_overview} after detrending with a second-order polynomial over time, presented in a chronological order according to their observing date. The residuals are shown in the right panel.}
         \label{plot_lcs}
   \end{figure*}

\subsection{Observations of the monitoring program}

We monitored HAT-P-12 with STELLA/WiFSIP in 2012 and in 2014. In 2012, the observations spanned 112 from April to August. If weather, target visibility, and scheduling constraints allowed for it, a sequence of seven exposures in Johnson V and seven exposures in Johnson I was obtained every night. On a few occasions, the robotic telescope system observed multiple 7+7 blocks per night separated by a few hours. In total, we gathered 636 images in V and 631 images in I. The exposure time was 30 and 60 seconds, respectively.

In 2014, the monitoring program was continued from February until July. At that point, 1081 images were taken in V, and 1077 images in I. The exposure time was 60 seconds in V and 50 seconds in I. The images were observed in blocks of five exposures per filter. In the year 2013, the CCD detector of WiFSIP was replaced, therefore the time series in 2012 and 2014 were not obtained with identical instruments. Additional telescope manufacturing work was done in May 2014, followed by the development of a new pointing model, which manifests in the time series as an abrupt change in detector position on June 2. Unfortunately, this change was accompanied by a jump in the differential light curve of about one percent amplitude.


\section{Data reduction}

\subsection{Transit data}
The reductions of the transit data obtained on STELLA, LBT, WHT, TNG, NOT, and OAdM were done homogeneously with a customized ESO-MIDAS pipeline, which calls the photometry software \textit{SExtractor} \citep{Bertin96}. Bias and flat-field correction was done in the usual way, with the bias value extracted from the overscan regions. Cosmic ray correction was carried out within SExtractor. We performed aperture photometry with circular apertures (MAG\_APER in SExtractor) and automatically adjusted elliptical apertures (MAG\_AUTO in SExtractor). The elliptical aperture offers the advantage of accounting for variations in the PSF over time, but with the disadvantage of different apertures for target and comparison stars. 

Our algorithm automatically searches for the combination of comparison stars, which minimizes the rms of the target light curve. The comparison stars are weighted according to their individual light curve quality in terms of point-to-point scatter (rms) \citep{Broeg2005}. The rms of the target light curve is computed not
only on the out-of-transit part, but also on the residuals of the entire time series after subtracting a transit fit including a second-order polynomial in time for detrending (see Section \ref{chap_anal}). As expected, the algorithm tends to choose only a low number of nearby stars in non-photometric conditions because atmospheric extinction differs for stars spatially separated on the sky. In contrast, in photometric conditions, the number of chosen comparison stars is higher (typically 5 to 8), including stars with larger spatial distance.

In a last step, the algorithm repeats the light curve extraction many times with different aperture sizes to search for the aperture that minimizes the scatter in the differential target light curve. In the majority of our observations, the circular aperture with fixed radius (MAG\_APER) yielded the more stable photometry than the flexible elliptical aperture (MAG\_AUTO).

The LBT transit observation suffered from highly variable sky transparency due to clouds moving through the FoV. We restrict the time series to data points that reached at least half the maximum flux value. We carried out a fringe correction for the LBC F972 light curve that resulted in a marginal depression of the point-to-point scatter of the final transit light curve from 1.01 mmag to 0.98 mmag. The values of the transit parameters remained unchanged.

The Asiago data were reduced with the software package STARSKY, described in detail in \cite{Nascimbeni2013a}. Figure \ref{plot_lcs} shows the differential light curves of all transits after normalization and detrending with a second-order polynomial over time.

\subsection{Monitoring data}
For the monitoring data, the bias and flatfield correction was done with the STELLA data reduction pipeline (for details see Granzer et al. 2015, in preparation). The following reduction steps were done with routines written in ESO-MIDAS. We conducted aperture photometry with SExtractor applying the MAG\_AUTO option to automatically adjust elliptic apertures. This method provides the flexibility to account for the varying observing conditions over the ten months of observing time. The I band data were fringe-corrected. We verified that the three used comparison stars GSC2.2-N130301273, GSC2.2-N130301284, and GSC2.2-N130301294 were photometrically stable to our achieved precision.

We discarded data points of very high airmass, low flux caused by poor weather, extremely high sky background (twilight), extreme values of FWHM indicating problems with the telescope focus, and extreme PSF elongation that indicates vibrations of the telescope caused by wind. Furthermore, we excluded the data taken after June 2, 2014 with the new pointing model, as the new detector position caused a jump in the differential light curve of about 1.0\% in V and 0.5\% in I. Additionally, we applied a 5\,$\sigma$ clipping to remove persistent outliers. The final light curve in V contains 486 of the 636 observed data points in 2012 and 873 of the 1081 data points in 2014. The I band light curve contains 402 out of 631 data points for 2012 and 709 out of 1077 data points for 2014.

Finally, the V band light curve of 2012 has a length of 109~days with a scatter of 5.1~mmag (rms), while the data points of 2014 cover 112~days and show a deviation of 3.2~mmag. The deviation is lower simply because the exposure time was doubled in 2014. The I band light curve from 2012 scatters by 3.0~mmag, which is comparable to the scatter of 2.6~mmag of the I band data from 2014.

\section{Transit modeling and results}
\label{chap_anal}

\subsection{Global transit parameters}
Throughout this work, we model the transit light curves with the publicly available software \mbox{JKTEBOP}\footnote{http://www.astro.keele.ac.uk/jkt/codes/jktebop.html} \citep{Southworth04}. It allows for a simultaneous fit for the transit model and detrending. Most ground-based differential light curves show a long-term trend potentially caused by atmospheric effects due to a color difference between the target and the combined reference stars. Throughout this HAT-P-12b analysis, we use a second-order polynomial over time to detrend the light curves. The transit fit parameters consist of the sum of the fractional planetary and stellar radius, $r_{\star} + r_p$, and their ratio $k=r_p/r_{\star}$, the orbital inclination $i$, the transit midtime $T_0$, and the host-star limb-darkening coefficients (LDC) $u$ and $v$ of the quadratic limb darkening law. The index ``$\star$'' refers to the host star and ``p'' refers to the planet. The dimensionless fractional radius is the absolute radius in units of the 
orbital semi-major axis $a$, $r_{\star} = R_{\star}/a$, and $r_p = R_p/a$. The planetary eccentricity is fixed to zero following \cite{Hartman2009} and the orbital period $P_{\mathrm{orb}}$ to 3.21306 days according to \cite{Lee2012}. Additional fit parameters are the three coefficients $c_{0,1,2}$ of the parabola over time.  

Throughout this work, we model the stellar limb darkening with a quadratic limb darkening law. Initial values for the LDC $u$ and $v$ are found by interpolating tabulated ATLAS values from \cite{Claret2011} for the stellar parameters of HAT-P-12: $T\,=\,4650$~K, log $g = 4.61$, [Fe/H]$ = -0.29$, and $v_{mic} = 0.85$~km s$^{-1}$ \citep{Hartman2009}. In the transit modeling, we fit for the linear coefficient $u$ but fix $v$ to its theoretical value. This ensures enough flexibility that the transit fits are not significantly biased by systematic errors of the LDC calculation. On the other hand, it avoids parameter degeneracies that occur when fitting for both $u$ and $v$. We account for errors in the calculated $v$ by a perturbation of this value by $\pm0.1$ during the error estimation of the transit parameters. 

For the error estimation, we run a Monte Carlo simulation (``task 8'' in JKTEBOP) of 5000 steps and the residual-permutation algorithm \citep[``task 9'',][]{Jenskins2002,Southworth08} and choose the higher value. The latter method takes the presence of correlated noise into account.

\subsubsection{Fit of individual transits}
\label{chap_indfits}

In a first step we apply a 4 $\sigma$ rejection to the light curve residuals after subtracting an initial model, with $\sigma$ as the standard deviation. A subsequent fit is used to rescale the individual photometric error bars: The error bars supplied by SExtractor yield a reduced $\chi^2$ that is always slightly greater than 1, which indicates that the photometric error bars are underestimated. Therefore, we enlarge the error bars by a factor individual to each light curve to give a $\chi^2_\mathrm{red}$ of unity. Furthermore, we calculate the so-called $\beta$ factor, a concept introduced by \cite{Gillon06} and \cite{Winn08} to include the contribution of correlated noise in the light curve analysis. It describes the evolution of the standard deviation $\sigma $ of the light curve residuals when they become binned in comparison to Poisson noise. In the presence of correlated noise, $\sigma_N $ of the binned residuals is larger by the factor $\beta$ than with pure uncorrelated (white) noise. There is a 
dependence of this value on the bin size, and generally the duration of ingress/egress is used as a relevant time scale for transit photometry \citep[e.g.,][]{Winn08}. We estimate the $\beta$ factor given in Tables \ref{tab_overview} and \ref{tab_litdata} as the average of the values from 10- to 25-minute bins, since the ingress of HAT-P-12b lasts about 19 minutes. It may happen that the derived $\beta$ factor is slightly smaller than unity because of statistical fluctuations. Such cases were manually set to $\beta = 1.0 $. The derived $\beta$ factor for each transit light curve is used to further enlarge the individual photometric error bars, and the final transit fit was performed including error estimation.

We can compare the scatter of the transit parameters derived per light curve with their one-sigma uncertainties to draw a rough conclusion about the reliability of the uncertainties: The 25 set of parameters from the complete transits analyzed in this work give a reduced $\chi^2 $ for the inclination $i$ of 2.2, which can be interpreted as slightly underestimated error bars. For the fractional stellar radius $r_{\star}$, the reduced $\chi^2 $ is 1.0 and 0.9 for the linear LDC $u$ (derived values of $u$ were compared with their theoretical predictions), an indication for error bars that match the scatter within these 25 measurements very well. The reduced $\chi^2 $ for the radius ratio $k$ is rather high with a value of 3.5, so that either the errors are underestimated, or else we see intrinsic effects such as a potential wavelength dependence (see Section \ref{chap_H12_transm}) or light curve modifications caused by starspots (see Section \ref{chap_H12_monit}), for which 
$k$ is the most sensitive transit parameter.

\subsubsection{Re-analysis of the literature data}
\label{reana_lit}

\begin{table}
\tiny
\caption{Overview of published transit observations of HAT-P-12b, which are re-analyzed in this work.}
\label{tab_litdata}
\begin{center}
\begin{tabular}{llcccr}
\hline
\hline
\noalign{\smallskip}

Date     &  Filter &  $N_{\mathrm{data}}$ &  rms    &  $\beta$ &  Reference \\
         &         &                      &  (mmag) &          &             \\
\hline
\noalign{\smallskip}

2007 Mar 27  &   i'     &     151  & 1.78  &   1.00   &   \cite{Hartman2009}  \\
2007 Apr 25  &   z'     &     375  & 2.62  &   1.02   &   \cite{Hartman2009}  \\
2009 Feb 5   &   z'     &     219  & 2.66  &   1.04   &   \cite{Hartman2009}  \\
2009 Mar 6   &   g'     &     214  & 5.14  &   1.31   &   \cite{Hartman2009}  \\
2010 May 31  &   J      &     383  & 2.52  &   1.33   &   \cite{Sada2012}     \\
2011 Mar 29  &   R      &     224  & 1.75  &   2.24   &   \cite{Lee2012}      \\
2011 Apr 14  &   R      &     273  & 2.64  &   1.68   &   \cite{Lee2012}      \\
2011 May 13  &   R      &     314  & 1.80  &   1.57   &   \cite{Lee2012}      \\
\hline
\end{tabular}
\end{center}
\end{table}

We analyze the published transit light curves described in Table \ref{tab_litdata} again. Before we include the literature light curves in our data sample, we test whether our analysis routines are able to reproduce the published results. Whenever possible we treat the published data identically to our observations (see Section \ref{chap_indfits}). For the \cite{Hartman2009} light curves, we achieve excellent agreement in the parameters, but find error bars that are higher by up to a factor of 2. The \cite{Lee2012} results, which were also derived with JKTEBOP, are reproduced as well, however our error estimation yields values higher by a factor of 2 for the planet-to-star radius ratio and 10 for the orbital inclination. The source of this discrepancy remains unexplained. The J-band light curve from \cite{Sada2012} shows rather complex deviations from the symmetric transit shape and was originally detrended by a fifth-order polynomial over time fitted to the out-of-transit data. The uncertainty of this 
polynomial 
was not included in the transit parameter error estimation. Following this approach, we can reproduce the transit parameters to within 1\,$\sigma$ with about the same error bars. However, a simultaneous fit of detrending polynomial and transit model changes the parameters by up to 1.5\,$\sigma$ and increases the uncertainty by up to a factor of 2.

\subsubsection{Fit of multiple transits}

We next fit all light curves of one filter band simultaneously, again using the inflated photometric error bars of Section \ref{chap_indfits}. JKTEBOP does allow the simultaneous fit of multiple transit observations if one joint limb-darkening function can be applied. The software cannot attribute different LDC to subsamples of the input data, which prohibits the simultaneous fit of multicolor data. However, the current version 34 of JKTEBOP is able to assign independent detrending polynomials to the individual light curves, so the detrending is included in the transit fit. Free parameters of the fits are $r_p + r_{\star}$, $k$, $i$, $T_0$, the period $P_{\mathrm{orb}}$, the linear coefficient $u$ of the quadratic law, and a set of coefficients $c_{0,1,2} $ of the detrending polynomial for each individual transit observation. The quadratic LDC $v$ is fixed to its theoretical value, but perturbed during the error estimation.  For the B band transit, the only filter with a single transit measurement in 
Table \ref{transitparam_H12}, we also keep $u$ to its theoretical value to ensure that the fit converges at reasonable parameter values. Table \ref{transitparam_H12} lists the transit parameters derived from all complete transits of each band fitted simultaneously. For a simultaneous fit, it is necessary to convert all barycentric Julian Date (BJD) timings of the observations to a common time standard which is in this work the barycentric dynamical time (TDB) following the recommendation of \cite{Eastman2010}. The u' band transits are not listed because their rather low accuracy would not improve the final results. The final transit parameter values are formed as the weighted average of the seven filters. 

For all our filter bands with multiple transits, the derived uncertainties of the simultaneous fit are larger and more conservative than the uncertainties derived by a weighted average of the individually fitted transits ($\sigma^2\,=\,1/ \Sigma^n_{i=1} \sigma^{-2}_i $). A plausible explanation is that the error estimation becomes more sensitive to correlated noise when analyzing multiple transits simultaneously. This is indicated by the fact that the residual-permutation method gives larger uncertainties than the Monte Carlo simulation. However, the uncertainties in the seven bands seem too conservative in part: the final values for i and $r_{\star}$ yield a $\chi^2_{\rm{red}}$ of 0.2 and 0.1, respectively. For u, we derive a $\chi^2_{\rm{red}}$ of 1.0. Finally, the $\chi^2_{\rm{red}}$ for k decreases from 3.5 for the 25 individual complete transits to 1.3 for the seven per-band values.

Our final transit parameters are generally in good agreement with previous publications on HAT-P-12b. Most values agree within the 1\,$\sigma$ levels, and the strongest deviation of about 2\,$\sigma$ is found for $k$ between our value and the one from \cite{Hartman2009}. The rather high value for $i$ of \cite{Lee2012} matches the value of this work to 1\,$\sigma$ if we consider the uncertainty of $\pm 1.0$ deg, which we derive from a re-analysis of their light curves.

\begin{table*}
\caption{Transit parameters of HAT-P-12b derived in this work and comparison to previously published values. The final values were formed as the weighted average of the seven filters.}
\label{transitparam_H12}
\begin{center}
\begin{tabular}{lccccr}
\hline
\hline
\noalign{\smallskip}
Filter      &  $r_{\star}\,=\,R_{\star}/a$ &  $k\,=\,r_p/r_{\star}$ &  i (deg) & linear LDC $u$ & quad. LDC $v$ \\
\hline
\noalign{\smallskip}
B    & 0.08526 $\pm$ 0.00382  &  0.13916 $\pm$ 0.00311  & 89.00 $\pm$ 1.04  &  0.9150\tablefootmark{a}  & -0.0592     \\
g'   & 0.08584 $\pm$ 0.00251  &  0.13847 $\pm$ 0.00215  & 88.60 $\pm$ 0.48  &  0.8080 $\pm$ 0.0371 & -0.0122    \\
V    & 0.08510 $\pm$ 0.00602  &  0.13679 $\pm$ 0.00439  & 89.41 $\pm$ 1.24  &  0.7458 $\pm$ 0.0972 & 0.0961    \\
r'   & 0.08456 $\pm$ 0.00145  &  0.13611 $\pm$ 0.00150  & 89.06 $\pm$ 0.78  &  0.5829 $\pm$ 0.0299 & 0.1665   \\
R    & 0.08565 $\pm$ 0.00155  &  0.13826 $\pm$ 0.00167  & 89.07 $\pm$ 0.66  &  0.5790 $\pm$ 0.0377 & 0.1823    \\
i'   & 0.08551 $\pm$ 0.00282  &  0.13515 $\pm$ 0.00230  & 89.26 $\pm$ 1.01  &  0.5081 $\pm$ 0.0603 & 0.2198     \\
z'   & 0.08561 $\pm$ 0.00290  &  0.14082 $\pm$ 0.00204  & 88.92 $\pm$ 0.88  &  0.3079 $\pm$ 0.0590 & 0.2426     \\
\hline
\hline
\noalign{\smallskip}
\textbf{Final} &  \textbf{0.08532} $\pm$ \textbf{0.00085} & \textbf{0.13779} $\pm$ \textbf{0.00079} & \textbf{88.98} $\pm$ \textbf{0.29} & & \\
\hline
\noalign{\smallskip}
\cite{Hartman2009} &  0.0850 $\pm$ 0.0013    &  0.1406 $\pm$ 0.0013  & 89.0 $\pm$ 0.4 & & \\
\cite{Lee2012}     &  0.0852 $\pm$ 0.0012    &  0.1370 $\pm$ 0.0019  & 89.9 $\pm$ 0.1  & &   \\
\cite{Sada2012}    &  0.0891 $\pm$ 0.0045    &  0.1404 $\pm$ 0.0026  & 88.5 $\pm$ 1.0  & &  \\
\cite{Line2013}    &  0.0862 $\pm$ 0.0029    &  0.137 $\pm$ 0.0011   &  88.7 $\pm$ 0.6  & &  \\

\hline
\end{tabular}
\end{center}
\tablefoot{
\tablefoottext{a}{Both limb darkening coefficients are fixed to its theoretical values for filter B, see text for details.}
}
\end{table*}

\subsection{Ephemeris and O-C diagram}
\label{ephemeris_H12}
To derive the ephemeris and investigate potential transit timing variations, we fit all individual transits, complete or partial, with a transit model fixing $r_p + r_{\star}$, $k$, and $i$ to the final parameter values of Table \ref{transitparam_H12}. Free parameters are the transit midtime $T_0$, the linear LDC $u$ and the three coefficients of the detrending polynomial $c_{0,1,2} $. The uncertainties are estimated using a Monte Carlo simulation with 5000 steps. Although the light curves from \cite{Hartman2009} and \cite{Lee2012} used slightly different detrending functions, we approximate their error contribution by including the same three free parameters $c_{0,1,2} $ as for the newly observed light curves of this work. The fit and MC error estimation of the Sada J band light curve additionally contain the free parameters $c_{3}$, $c_{4}$, and $c_{5}$ to model the trend with a fifth-order polynomial over time. The quadratic LDC $v$ is fixed to its theoretical value. We do not perturb $v$ in the MC 
simulation since we find that $T_0$ is insensitive to the treatment of limb darkening.

Table \ref{tab_timing} summarizes all individual transit timings with their uncertainties and includes the individual deviation $O - C$ of the observation from the calculated ephemeris. All times are given in BJD(TDB). We use a linear least-squares fit to the transit times of the 26 complete transits to newly determine a linear ephemeris:

\begin{equation}
T_c\,=\,\mathrm{BJD(TDB)}\,2456032.151332(39)\,+\,3.21305756(20)\,N ,
\label{equ_ephem_H12}
\end{equation}

\noindent where $T_c$ is the predicted central time of a transit, $N$  the cycle number with respect to the reference midtime, and the numbers in brackets give the uncertainties of the last two digits estimated with a Monte Carlo simulation of 5000 steps. The reference midtime is chosen to minimize the covariance between reference midtime and period. The estimated orbital period differs by about 1\,$\sigma$ from the original value given in the discovery paper \cite{Hartman2009}, but is an order of magnitude more accurate. However, our period is 0.15 seconds shorter than the period value of \cite{Lee2012}, which is a deviation of about 5\,$\sigma$. We note that some of the very first transits of the data sample, the three complete Hartman transits, happen systematically earlier by about 65 seconds in the calculation of \cite{Lee2012} than in our calculation. Since the data of \cite{Hartman2009} are given in the time standard UTC, this deviation might be explained by a faulty UTC-TDB correction.

Simultaneously observed transits offer the possibility of testing the reliability of the derived transit timing uncertainties, although one needs to be careful to draw definite conclusions from such low number statistics. We observed the transit of March 15, 2014 with three different telescopes, the William Herschel Telescope observed with ULTRACAM in three different bands. We therefore have five transit light curves of the very same event. The reduced $\chi^2 $ value of these five transit midtimes is 2.32. We ran a simulation of 10~000 datasets of five points each, added Gaussian noise, and find that such a $\chi^2_{\rm{red}} $ or higher is reached in only 10\% of all cases. Therefore, it appears likely that the uncertainties are slightly underestimated. However, with a hypothetical increase of only 10\% on the error bars, the $\chi^2_{\rm{red}} $ value would already lie within 1\,$\sigma$ of the mean of the simulated $\chi^2_{\rm{red}}$ distribution.

\begin{table}
\tiny
\caption{Observed transit times of HAT-P-12b. }
\label{tab_timing}
\begin{center}
\begin{tabular}{lcrcl}
\hline
\hline
\noalign{\smallskip}
BJD(TDB) &  Uncertainty & Cycle & O - C & Reference \\
(2,450,000+) &     & number  &     &      \\
\hline
\noalign{\smallskip}
4187.85647\tablefootmark{a} &  0.00061  &  -574  &  0.00017  & \cite{Hartman2009} \\
4216.77343\tablefootmark{a} &  0.00023  &  -565  & -0.00037  & \cite{Hartman2009} \\
4869.02462\tablefootmark{a} &  0.00033  &  -362  &  0.00013  & \cite{Hartman2009} \\
4897.94297\tablefootmark{a} &  0.00057  &  -353  &  0.00095  & \cite{Hartman2009} \\
5347.76952\tablefootmark{a} &  0.00035  &  -213  & -0.00054  & \cite{Sada2012} \\
5630.51929 &  0.00011 &  -125 &  0.00015 & this work (Asiago) \\
5649.79770\tablefootmark{a} &  0.00035 &  -119 &  0.00022 &  \cite{Lee2012}  \\ 
5659.43649 &  0.00044 &  -116 & -0.00016 & this work (STELLA) \\
5665.86227\tablefootmark{a} &  0.00037 &  -114 & -0.00049 & \cite{Lee2012}  \\
5675.50436 &  0.00047 &  -111 &  0.00242 & this work (STELLA) \\
5691.56661 &  0.00022 &  -106 & -0.00061 & this work (STELLA) \\
5694.78089\tablefootmark{a} &   0.00029 &  -105 &  0.00060 & \cite{Lee2012} \\  
5707.63244  &  0.00027  &  -101  & -0.00006  & this work (STELLA) \\
5720.48443  &  0.00015  &   -97  & -0.00031  & this work (STELLA) \\
5736.54999  &  0.00030  &   -92  & -0.00004  & this work (STELLA) \\
5964.67787  &  0.00059  &   -21  &  0.00075  & this work (STELLA) \\
6006.44779  &  0.00012  &    -8  &  0.00091  & this work (Asiago) \\
6051.42937  &  0.00048  &     6  & -0.00030  & this work (STELLA) \\
6054.64202  &  0.00060  &     7  & -0.00070  & this work (STELLA) \\
6083.55936  &  0.00065  &    16  & -0.00088  & this work (STELLA) \\
6388.80101  &  0.00081  &   111  &  0.00029  & this work (LBT) \\
6388.80064  &  0.00048  &   111  & -0.00007  & this work (LBT) \\
6716.53165  &  0.00033  &   213  & -0.00093  & this work (STELLA) \\
6732.59827  &  0.00063  &   218  &  0.00039  & this work (STELLA) \\
6732.59765  &  0.00011  &   218  & -0.00022  & this work (NOT) \\
6732.59811  &  0.00028  &   218  &  0.00023  & this work (WHT) \\
6732.59741  &  0.00023  &   218  & -0.00046  & this work (WHT) \\
6732.59796  &  0.00009  &   218  &  0.00008  & this work (WHT) \\
6793.64833  &  0.00092  &   237  &  0.00236  & this work (OAdM) \\
6806.49642  &  0.00062  &   241  & -0.00178  & this work (STELLA) \\
6806.49815  &  0.00019  &   241  & -0.00004  & this work (TNG) \\
\hline
\end{tabular}
\end{center}
\tablefoot{
\tablefoottext{a}{Transit midtime newly estimated in this work.}
}
\end{table}

\begin{figure}
   \centering
   \includegraphics[height=\hsize,angle=270]{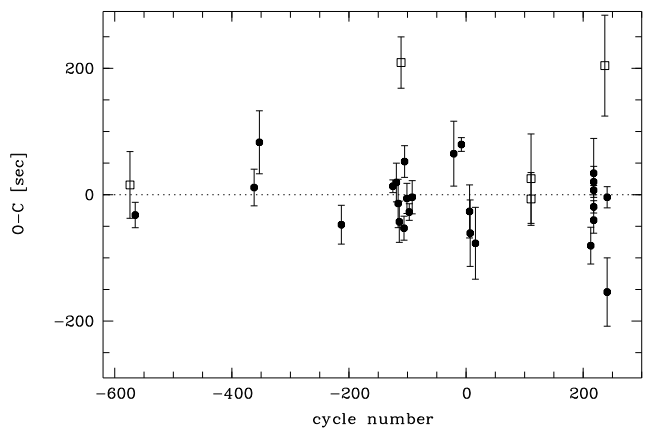}
      \caption{Transit timing residuals versus the linear ephemeris of Equation \ref{equ_ephem_H12}. The filled symbols represent complete transits, and the open squares indicate transits which were just partially observed.}
         \label{plot_ephe_H12}
   \end{figure}

Figure \ref{plot_ephe_H12} shows the deviations of the individual transit times from the linear ephemeris of Equation \ref{equ_ephem_H12}. All data points give $\chi^2_{\rm{red}} \sim 4.8 $, and an exclusion of the partial transit lowers this value only marginally to $\chi^2_{\rm{red}}\sim 4.5 $. The most deviating point is seven sigma away from the predicted ephemeris and corresponds to the March 19, 2012 transit from the Asiago telescope. If this point is excluded, the $\chi^2_{\rm{red}}$ lowers to 2.4.

\begin{figure}
   \centering
   \includegraphics[height=\hsize,angle=270]{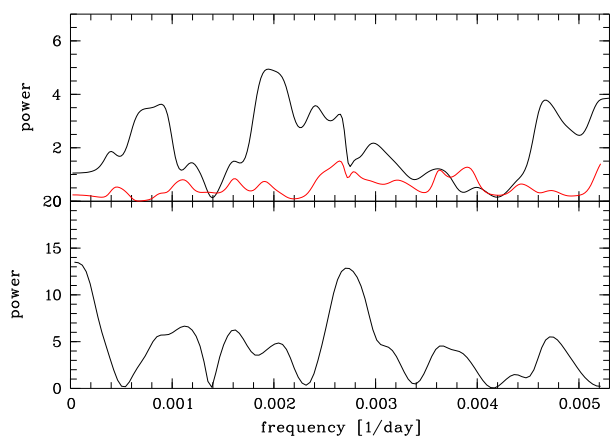}
      \caption{Upper panel: Lomb-Scargle periodogram of the timing deviation from the linear ephemeris (O-C) when excluding the partial transits (black). A marginal signal can be seen at a frequency of $\sim$0.0019 day$^{-1}$. In comparison, the periodogram without the transit of March 19, 2012 (epoch=-8) is also shown (red). Lower panel: Window function of the data set.}
         \label{plot_periodoTTV_H12}
   \end{figure}

We calculate a Lomb-Scargle periodogram of all transits times excluding the partial transits shown in Figure \ref{plot_periodoTTV_H12}. For a test, we again exclude the Asiago transit from March 19, 2012 and find a very different periodogram without any significant peak. The limit for the highest frequency is defined by the average Nyquist frequency. We conclude that we find no evidence of any transit timing variation and that the rather high value of $\chi^2_{\rm{red}} \sim 2.4$ is caused by an underestimation of the error bars due to systematics. One potential source of systematics in the light curves that influences the timing are starspots \citep{Barros2013}, a plausible scenario for this slightly variable host star (see Section \ref{chap_H12_monit}).

\subsection{Planetary radius over wavelength}
\label{chap_H12_transm}

One major goal of this work is to search for a wavelength dependence for the effective planetary radius. For this purpose, we fit all light curves again individually, this time by fixing the inclination $i$ and the fractional stellar radius $r_{\star}$ to the values derived in Section \ref{chap_anal} and the orbital period and transit midtime to values derived from the ephemeris (Eq. \ref{equ_ephem_H12}) in Section \ref{ephemeris_H12}. All these parameters are assumed to be wavelength independent, but their values have uncertainties. Here, we neglect $\Delta r_{\star} $, $\Delta i $, and $\Delta T_c $ under the assumption that they are a source of uncertainty common to all band passes following, for example, \cite{Mancini2014b}. The derived error bars $\Delta k $ are then relative errors. Free fit parameters of the model are 
the fractional planetary radius $r_p$ and the three coefficients of the detrending parabola $c_{0,1,2}$. The J-band light curve needs the additional fit parameters $c_{3,4,5}$ 
because of its more complex trend. For comparison we fit each transit twice, first with the linear stellar LDC $u$ as fit parameter and second with $u$ fixed to its theoretical value. The quadratic stellar LDC $v$ is fixed to its theoretical value in both cases but perturbed in the error estimations. The partial transit of the LBC covers too little orbital phase to reasonably constrain detrending and limb darkening in the same fit. Therefore, for the u' band light curve, $u$ is fixed to the value found in a simultaneous TNG + WHT u' band fit, and for the F972N20 filter, $u$ is fixed to the value derived by the z' band light curves.

\begin{table*}
\caption{Planet-to-star radius ratio $k$ per transit observation and per filter with relative uncertainties. The parameters $r_{\star}$, $i$, $T_0$, and $P_{\mathrm{orb}}$ are fixed to their values derived in Section \ref{chap_anal}. In Column 3, $k$ is calculated with $u$ as free parameter and $v$ fixed to theoretical value, but perturbed in error estimation. In Column 5, $k$ is calculated with $u$ and $v$ fixed to theoretical values.}
\label{k_indtrans_H12}
\begin{center}
\begin{tabular}{l|c|cc|cc|c}
\hline
\hline
\noalign{\smallskip}
   Date  & Filter &      \multicolumn{2}{c|}{ $u$ free }      &  \multicolumn{2}{c|}{ $u$ fixed }      &  $v$      \\
         &        &   $k$         &  $u$             & $k$ & $u$   &   fixed   \\
\hline
\noalign{\smallskip}
  \multicolumn{3}{l}{ Individual light curves: } & & & & \\
\hline
\noalign{\smallskip}
2013 Mar 5    &   u'   & 0.13874 $\pm$ 0.00216  &  1.0968\tablefootmark{a}             &  0.13947  &  1.0698 &  -0.2218  \\
2014 Mar 15   &   u'   & 0.13989 $\pm$ 0.00282  &  0.9922 $\pm$ 0.0797  &  0.13793  &  1.0698 &  -0.2218  \\
2014 May 28   &   u'   & 0.13627 $\pm$ 0.00263  &  1.1778 $\pm$ 0.0741  &  0.13914  &  1.0698 &  -0.2218  \\
2014 Mar 15   &   B    & 0.13836 $\pm$ 0.00106  &  0.9428 $\pm$ 0.0301  &  0.13905  &  0.9154 &  -0.0592  \\
2014 Feb 27   &   g'   & 0.13932 $\pm$ 0.00119  &  0.8046 $\pm$ 0.0346  &  0.13879  &  0.8492 &  -0.0122  \\
2014 Mar 15   &   g'   & 0.13328 $\pm$ 0.00125  &  0.9074 $\pm$ 0.0359  &  0.13392  &  0.8492 &  -0.0122  \\
2014 Mar 15   &   g'   & 0.13605 $\pm$ 0.00049  &  0.8662 $\pm$ 0.0143  &  0.13620  &  0.8492 &  -0.0122  \\
2012 Feb 6    &   V    & 0.14103 $\pm$ 0.00244  &  0.7060 $\pm$ 0.0798  &  0.14180  &  0.6934 &   0.0961  \\
2012 May 3    &   V    & 0.13615 $\pm$ 0.00179  &  0.6776 $\pm$ 0.0607  &  0.13661  &  0.6934 &   0.0961  \\
2012 May 6    &   V    & 0.13821 $\pm$ 0.00221  &  0.8590 $\pm$ 0.0695  &  0.14114  &  0.6934 &   0.0961  \\
2012 Jun 4    &   V    & 0.13783 $\pm$ 0.00251  &  0.6575 $\pm$ 0.0857  &  0.13785  &  0.6934 &   0.0961  \\
2011 May 9    &   r'   & 0.13915 $\pm$ 0.00097  &  0.5350 $\pm$ 0.0370  &  0.13881  &  0.5831 &   0.1665  \\
2011 May 25   &   r'   & 0.13951 $\pm$ 0.00103  &  0.6473 $\pm$ 0.0659  &  0.13923  &  0.5831 &   0.1665  \\
2011 Jun 7    &   r'   & 0.13731 $\pm$ 0.00102  &  0.5796 $\pm$ 0.0397  &  0.13749  &  0.5831 &   0.1665  \\
2011 Jun 23   &   r'   & 0.14187 $\pm$ 0.00122  &  0.5818 $\pm$ 0.0416  &  0.14164  &  0.5831 &   0.1665  \\
2014 Mar 15   &   r'   & 0.13377 $\pm$ 0.00038  &  0.6047 $\pm$ 0.0129  &  0.13420  &  0.5831 &   0.1665  \\
2011 Mar 29   &   R    & 0.13930 $\pm$ 0.00152  &  0.7328 $\pm$ 0.0484  &  0.14312  &  0.5450 &   0.1823  \\
2011 Apr 14   &   R    & 0.14176 $\pm$ 0.00160  &  0.6319 $\pm$ 0.0579  &  0.14325  &  0.5450 &   0.1823  \\
2011 May 13   &   R    & 0.14067 $\pm$ 0.00130  &  0.5322 $\pm$ 0.0767  &  0.14044  &  0.5450 &   0.1823  \\
2011 Mar 9    &   R    & 0.13880 $\pm$ 0.00064  &  0.5977 $\pm$ 0.0213  &  0.13984  &  0.5450 &   0.1823  \\
2012 Mar 19   &   R    & 0.13598 $\pm$ 0.00062  &  0.5806 $\pm$ 0.0161  &  0.13687  &  0.5450 &   0.1823  \\
2007 Mar 27   &   i'   & 0.13931 $\pm$ 0.00205  &  0.2295 $\pm$ 0.1119  &  0.14029  &  0.4379 &   0.2198  \\
2014 May 28   &   i'   & 0.13597 $\pm$ 0.00083  &  0.4786 $\pm$ 0.0310  &  0.13715  &  0.4379 &   0.2198  \\
2007 Apr 25   &   z'   & 0.14047 $\pm$ 0.00107  &  0.3261 $\pm$ 0.0381  &  0.14018  &  0.3518 &   0.2426  \\
2009 Feb 5    &   z'   & 0.14024 $\pm$ 0.00142  &  0.3633 $\pm$ 0.0532  &  0.14057  &  0.3518 &   0.2426  \\
2013 Apr 5    &   F972 & 0.13888 $\pm$ 0.00182  &  0.3302\tablefootmark{b}             &  0.13868  &  0.3518 &   0.2426  \\
2010 May 31   &   J    & 0.13775 $\pm$ 0.00285  &  0.3660 $\pm$ 0.0637  &  0.14033  &  0.2200 &   0.3120  \\
\hline
\noalign{\smallskip}
  \multicolumn{4}{l}{ Combination of light curves per filter: } & & & \\
\hline
\noalign{\smallskip}
         &  u'    & 0.13728 $\pm$ 0.00244  &  1.0968 $\pm$ 0.0656 &           & 1.0698  &  -0.2218  \\
         &  g'    & 0.13633 $\pm$ 0.00074  &  0.8711 $\pm$ 0.0123 &           & 0.8492  &  -0.0122  \\
         &  V     & 0.13857 $\pm$ 0.00131  &  0.7018 $\pm$ 0.0394 &           & 0.6934  &   0.0961  \\ 
         &  r'    & 0.13741 $\pm$ 0.00071  &  0.5833 $\pm$ 0.0130 &           & 0.5831  &   0.1665  \\
         &  R     & 0.13793 $\pm$ 0.00098  &  0.5870 $\pm$ 0.0201 &           & 0.5450  &   0.1823  \\ 
         &  i'    & 0.13703 $\pm$ 0.00103  &  0.4695 $\pm$ 0.0297 &           & 0.4379  &   0.2198  \\
         &  z'    & 0.14035 $\pm$ 0.00087  &  0.3302 $\pm$ 0.0289 &           & 0.3518  &   0.2426  \\
                                                         
\hline
\end{tabular}
\end{center}
\tablefoot{The uncertainties of $k$ with $u$ fixed to its theoretical value are on average 20 percent smaller than with $u$ as free to fit the parameter, but are less realistic because they do not include an uncertainty of $u$. For the combined fit of multiple transits per filter, the fitted values of $u$ generally agree with their theoretical values.
\tablefoottext{a}{The linear limb darkening coefficient $u$ was fixed but perturbed to the value derived from the u' band light curves 2014, Mar 15 and 2014, May 28.}
\tablefoottext{b}{The linear limb darkening coefficient $u$ was fixed but perturbed to the value derived from the z' band light curves 2007 Apr 25 and 2009 Feb 5.}
}
\end{table*}

   \begin{figure}
   \centering
   \includegraphics[height=\hsize,angle=270]{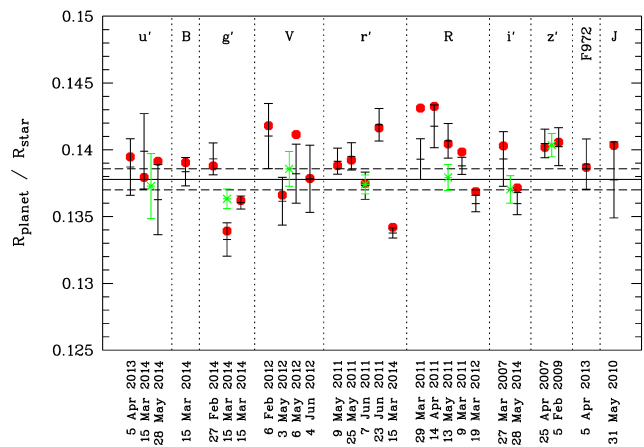}
      \caption{Planet-to-star radius ratio $k$ for the individual transit light curves. Black crosses show $k$ when the linear LDC $u$ is a free-fit parameter, and red points show the same value when $u$ is fixed to its theoretical value given in Table \ref{k_indtrans_H12}. The error bars of the latter are on average 20 percent smaller than the uncertainties of the former so are not shown here. Data points in green show the value of $k$ for the simultaneous fit of multiple transits per filter. The horizontal line shows the globally derived value of $k$ as given in Table \ref{transitparam_H12} and its uncertainty in dashed lines.}
         \label{plot_ind_k}
   \end{figure}

We combine the data sets per filter band again in a simultaneous fit. Free parameters are $r_p$, $u$, and a set of coefficients $c_{0,1,2} $ for every individual transit observation. The two partial STELLA r' transits and the partial OAdM transit are not analyzed here. In general, we note very good agreement of the observationally derived linear LDC per filter and their theoretical calculations. Only in the two filter sets mostly affected by red noise, R and J, do we see deviations of 2.1 and 2.3\,$\sigma$.

The results for individual transits and per-filter fits are given together in Table \ref{k_indtrans_H12} and Figure \ref{plot_ind_k}.

\section{Results of the monitoring program and starspot correction}
\label{chap_H12_monit}

\begin{figure*}
   \centering
   \includegraphics[height=13cm,angle=270]{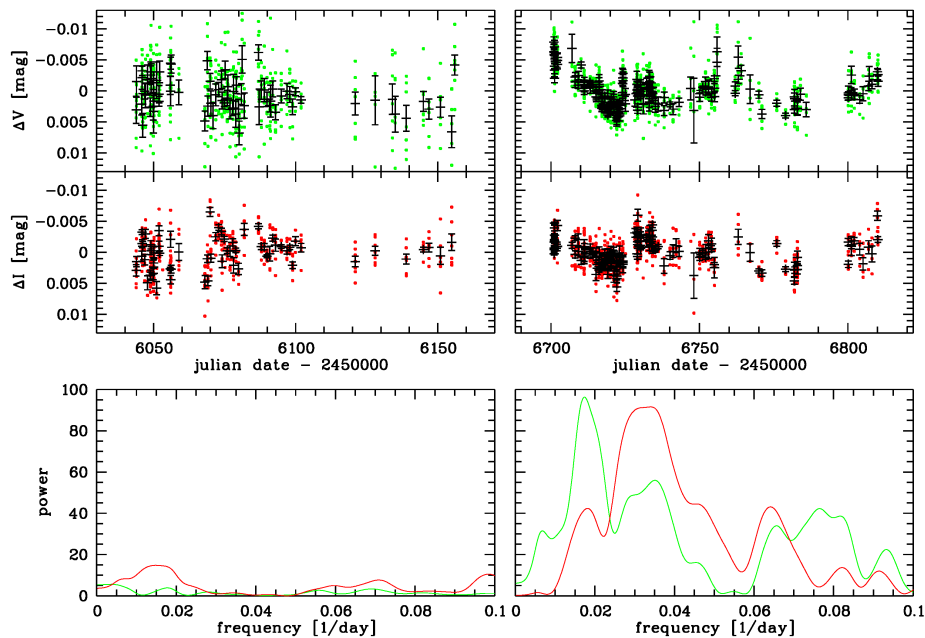}
      \caption{Upper panel: STELLA/WiFSIP differential V band photometry with the original data points in green and the average of the five exposures taken as a block given in black, with the error bars as the standard deviation divided by the square root of the number of averaged data points. On the left the observations from 2012 are shown, on the right the data from 2014. Middle panel: Same as in the upper panel, here for the I-band photometry. Lower panel: The Lomb-Scargle periodogram for the V band data (green) and the I-band data (red), on the left for 2012, on the right for 2014.}
         \label{plot_lotephot_H12}
   \end{figure*}

We monitored HAT-P-12 in 2012 and 2014 for about four months each in two colors, V and I, to investigate photometric variations potentially caused by starspots. Such brightness inhomogeneities on the stellar photosphere, whether occulted or unocculted by the transiting planet, modify the derived transit parameters, especially the transit depth \citep{Sing2011b,Pont2013,Oshagh2014}. The differential light curves are presented in Figure \ref{plot_lotephot_H12}. A calculated Lomb-Scargle periodogram for V and I in 2012 shows no compelling periodicity in this dataset. The periodogram of the V band data of 2014 shows indications of variations with a period of about 60~days. A fit of a sine curve with this period yields a semi-amplitude of $1.8 \pm 0.4$~mmag. The I-band data set of 2014 also shows mild indications for this period, but a period of about half this value, $P\sim 30$~days, is more significant. A sine fit yields an amplitude of $1.9 \pm 0.3$~mmag for the 30-day period and $1.4 \pm 0.4$~mmag 
for the period of 60~days. We calculated plain least-squares periodograms to estimate the amplitude of periodic sine and cosine functions over a wide period range and concluded that periodic signals of semi-amplitudes larger than about 2~mmag are excluded by the V and I data of 2014. For the observation period of 2012, we can exclude periodic signals of semi-amplitudes higher than about 1.5 mmag. 

An inspection of the light curves by eye reveals a strongest measured variation of almost 1\% in the V band data of Season 2014 between Julian date 2456701 and 2456723. The difference $\Delta f$ of the measured stellar flux and the flux of the unspotted photosphere might be larger than the measured variability, because a certain level of permanently visible starspots may exist. We approximate this level of permanent flux dimming by the variance of our monitoring light curve \citep{Pont2013} and obtain $\Delta f \sim 1.3$\%. If we assume a temperature contrast between photosphere and spots of 1000 K \citep{Strassmeier2009}, the filling factor would amount to $\sim$\,0.016. Correspondingly, the maximum correction for unocculted spots ranges from about $\Delta k = 0.0010$ in the u' band to about $\Delta k = 0.0006$ at J using Equation 4 in \cite{Sing2011b} and the assumption of black body radiation for the photosphere and the spots. These correction values resemble our 1\,$\sigma$ uncertainties for 
$k$ in Table \ref{k_indtrans_H12}. However, this estimation is based on the 1\% measured flux variation, and none of our measurements took place at this moment of highest measured activity. Instead, to all transits observed during the monitoring campaigns, we could attribute filling factors of only half the mentioned value or less. Therefore, the contribution from an unocculted star spot would be significantly smaller than our uncertainties in most cases, so we do not include this correction.

Although the correction for occulted spots could in principle be larger, we keep the uncorrected values in our analysis for the following reasons:

\begin{itemize}
 \item At the level of precision of our light curves, we do not see any indication of a bump caused by a starspot crossing.
 \item Transits affected mostly by occulted spots are shallower than the average value. The transits of 2014 March 15 show a shallower transit in g' and r'. However, their simultaneously taken equivalents in B and u' are not shallower than the average value. Because the effect of occulted spots would be enhanced for these bluer wavelengths, we conclude that the lower values of $k$ in g' and r' are caused by different kinds of systematics.
 \item If the spots are not homogeneously distributed along the transit chord, they cause deformations that is treated like red noise in our analysis, so that the uncertainty introduced by these spots is included in the uncertainty of the transit parameters. In the less likely case of spots evenly distributed along the transit chord (resulting in a symmetric transit shape without bumps), the shallower transit curve would not match other transits taken at moments of less stellar activity and be considered as correlated noise in the error analysis.
 \item The monitoring campaigns have shown that the activity is time dependent, with times of no measurable spot modulation. We interpret these as times with very low spot activity and assume that the simultaneous analysis of transits taken at different epochs weakens the influence of spots crossed at times of higher activity. 
\end{itemize}

\section{Discussion}

\subsection{Transmission spectrum}
\begin{figure}
   \centering
   \includegraphics[height=\hsize,angle=270]{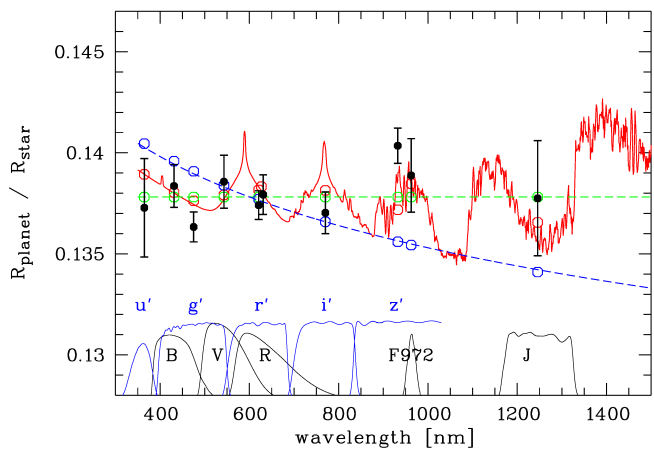}
      \caption{Broad-band transmission spectrum of HAT-P-12b. The measured values are given in black. Overplotted are a cloud-free solar metallicity spectrum of HAT-P-12b from \cite{Fortney2010} (red solid line), a Rayleigh-scattering slope $dR_p/d\mathrm{ln}\lambda\,=\,-4H$ (blue solid line), and a flat line presenting wavelength-independent atmospheric absorption (green solid line).  When folded with the filter, transmission curves, and a
theoretical spectrum of HAT-P-12, the model values are given in their corresponding color as open circles. At the bottom the transmission curves of the used filters are shown.}
         \label{transspec_H12}
   \end{figure}

We compare the measured transmission spectrum (referring to the common fits in each band) to three different theoretical atmosphere models: The first model presents a cloud-free atmosphere and was computed by \cite{Fortney2010} using a planet-wide average P-T profile, solar metallicity, and the planet parameters of HAT-P-12b. The second model simulates an atmosphere with a haze layer by the introduction of a Rayleigh-scattering slope of $dR_p/d\mathrm{ln}\lambda\,=\,-4H$ similar to the observed spectrum of HD189733b \citep{Leca2008b}. Here, $H$ is the atmospheric scale height and equals $\sim$\,600~km in the case of HAT-P-12b. The third model is a simple flat line that represents an atmosphere dominated by clouds that cause gray absorption. We multiply all models by the filter transmission curves and a theoretical spectrum of HAT-P-12 computed with the SPECTRUM spectral synthesis code \citep{Gray1994} to obtain theoretical values for k. The match between these three models and our data was computed using a $\
chi^2$ 
test. All three models were offset in the vertical direction until their mean from the B to the z' band equaled the global $k$ value of Table \ref{transitparam_H12}, $k=0.13779$, see Figure \ref{transspec_H12}. In this way, we set the pressure level that defines the planetary radius. The achieved values are $\chi^2_{\rm{red}} = 1.61$ for the flat line model, 2.20 for the cloud-free model, and 5.92 for the Rayleigh slope with the number of degrees of freedom equaling nine. The corresponding one-tailed probabilities $P$ of a $\chi^2$ test are 0.11, 0.019, and $<1\times10^{-5}$. This indicates that the measured data points agree with the flat model spectrum, marginally agree with the cloud-free model spectrum, but disagree with the Rayleigh slope. Moreover, the discrepancy of the g' and z' band measurements of 3.5\,$\sigma$ points toward a wavelength dependence of the planetary radius.

However, individual data points might still be affected by correlated noise. Therefore, we base our results on the ensemble of data points in which individual systematic effects are minimized. As a test, we rejected the most deviant data point (that of the z' band) and compared this new spectrum with the same three models. After a newly calculated vertical shift, the $\chi^2_{\rm{red}}$ value for the flat, cloud-free, and Rayleigh models are now 0.50, 0.62, and 2.73, respectively. Thus, the data sample is in very good agreement with the flat line, but cannot clearly distinguish between the constant planetary radius and the cloud-free model. However, the corresponding probability for the Rayleigh slope of $5\times10^{-3}$ still rules out this model after discarding the z' band. This probability rises slightly if we take into account that a constant Rayleigh slope is unlikely to be sustained over many scale heights owing to grain settling \citep{Pont2013}. In the case of 
HD189733b, the Rayleigh slope is 
constant from the UV to the z' band \citep{Sing2011b}, but weakens at redder wavelengths until $k$ reaches a constant value at about the H or K band \citep{Pont2013}. To simulate this effect, we adjusted the same value of $k$ for the J band as for the F972N20 filter for the Rayleigh model, which changes $\chi^2_{\rm{red}}$ to 2.45 and increases the probability to $1\times10^{-2}$. Fixing $u$ to its theoretical value reduces the agreement of the measurements with the Rayleigh slope model since all data points remain nearly unchanged except the point at J, which increases by about 1\,$\sigma$. A Rayleigh slope confined to a narrower wavelength range in the blue followed either by a flat line or a cloud-free spectrum at redder wavelengths cannot be ruled out by our data.

In a previous study, \cite{Line2013} obtained a transmission spectrum from 1.1 to 1.7~$\mu$m, which clearly rules out the presence of the broad water absorption feature at 1.4~$\mu$m predicted by a cloud-free atmosphere model. They conclude that clouds in the atmosphere are the physically most plausible explanation for their rather flat spectrum. Our work extends the observed wavelength coverage of HAT-P-12b toward the near UV and leaves the cloudy atmosphere as the best explanation for the entire spectrum.

HAT-P-12b has an equilibrium temperature of about 1000 K. Theoretical models of hot-Jupiter atmospheres predict cloud-free atmospheres as clouds of, for example, MgSiO$_3$ are expected to form at higher pressure levels than those probed by these observations \citep[see][and references therein]{Madhusudhan2014}. However, HAT-P-12b is not the only hot Jupiter of this temperature regime that does not fulfill the theoretical predictions. The works on transmission spectroscopy of \cite{Gibson2013} on WASP-29b, \cite{Nikolov2014} on HAT-P-1b, and \cite{Mallonn2015} on HAT-P-19b disfavor clear atmospheres. Although for each case there might be individual explanations including for example alkali metal depletion by photoionization or night-side condensation, one scenario eases the interpretation of all these observations: the slant viewing geometry of this observing technique through the planetary atmosphere strongly enhances the opacities of minor condensates. Even in low concentrations, such material can mask 
atmospheric 
spectral features \citep{Fortney2005}. Therefore, in the context of current instrumental limits, transmission spectra with muted features might outweigh the detections of broad cloud-free features.

\subsection{Periodicity in the monitoring light curves}
The V band light curve of season 2014 presented in Section \ref{chap_H12_monit} shows a periodicity of $\sim$\,60~days, which the I band one of $\sim$\,30~days, thus the question arises if these are linked to the stellar rotation period. The semi-amplitudes of the order 1-2~mmag are well within the regime of systematic effects. To verify this signal, we investigated independent photometric data sets. We downloaded the monitoring data of HAT-P-12 of the SuperWASP survey from the NASA Exoplanet Archive\footnote{exoplanetarchive.ipac.caltech.edu/docs/SuperWASPMission.html} observed in 2007 and found a periodicity of $\sim$\,65~days, here with a slightly larger amplitude of about 3~mmag. However, not far from this period, the window function of this data set has a moderate peak at $\sim$\,80~days. We also analyzed the discovery light curve from the HATnet survey, taken in 2006 and another HATnet monitoring of this object from the year 2010, both made available on the HATnet homepage\footnote{hatnet.org/planets/
discovery-
hatlcs.html}. Their photometric precision only rules out significant periods with amplitudes higher than about 4~mmag, but does not allow for any statement about the potential periods at 30 and 60~days. Further observations are needed to confirm that these potential periods can be attributed to the stellar rotation.

\section{Conclusion}
                                                                                                                                                                                                                                                                                         
The scientific goal of this work has been to investigate the transmission spectrum of HAT-P-12b for spectral variations from the near-UV (sloan u') to the near-IR (Johnson J) using broad-band filters. We observed 20 transit events with seven different facilities. Two events were observed in multichannel mode, thus in total we acquired 23 new photometric transit light curves. We also re-analyzed eight transit light curves already published in the literature to complete our filter sample, which finally consisted of ten different filters. In addition, the host star HAT-P-12 was monitored in the observing seasons 2012 and 2014 to evaluate its activity level and the level of starspot influence on the derived transit parameters.

The entire sample of transit light curves was homogeneously analyzed with the transit modeling software JKTEBOP. The derived transit parameters are generally in good agreement with previously published values, but with increased accuracy. Furthermore, the large number of light curves and their time separation (2007 -- 2014) allows for a refinement of the orbital ephemeris. No hint of a periodic transit timing variation was found. However, the scatter of the data points in the O\,--\,C diagram is slightly greater than their mean error bar. Further investigations are needed to verify whether the scatter is increased by underestimated correlated noise in the data or by an astrophysical effect.

The monitoring data show in general a photometrically quiet object. We measured a maximum variability of one percent with no periodicity. The data allowed us to exclude any periodic variations of an amplitude greater than 2~mmag for the monitored time span. We concluded that a potential influence of starspots on the transit parameters is within the derived transit parameter uncertainties, therefore we did not apply an additional starspot correction. Our monitoring campaign, together with the HATnet and SuperWASP photometry of HAT-P-12, yields indications of a photometric variation of 60-day period with less than 2~mmag amplitude, however the signal is not significant enough to be attributed to the stellar rotation.

In agreement with \cite{Lendl2013}, we found that multiple transit observations lower the effect of uncorrected correlated noise on the derived transit parameters. Because this noise component is not correlated between individual light curves, its effect is minimized when averaged. Furthermore, the error estimation results in more realistic error bars if multiple light curves are analyzed simultaneously, indicated here by the better agreement of the transit parameters derived after combining the light curves per filter compared to those extracted from the individual light curves.

The combined ten data points over wavelength, which form our transmission spectrum, were compared to a flat line, a Rayleigh scattering slope (power law coefficient $\alpha=4$), and a cloud-free solar-composition model. In general, the data favor the flat model. However, individual data points deviate significantly
from each other, especially the z' band, which shows a deeper transit. While additional observations need to confirm the reality of this measurement, a Rayleigh scattering model extending into the NIR can be robustly excluded independently of this individual filter. However, a Rayleigh slope confined to shorter wavelengths cannot be ruled out. Our spectrum strengthens the conclusion of \cite{Line2013} obtained in the NIR that gray absorption caused by clouds is the best explanation for the atmosphere of HAT-P-12b, and we extend this result to a broader wavelength range from the near-UV to the NIR.

%

\begin{acknowledgements}
Based on observations made with the Large Binocular Telescope (LBT), the Italian Telescopio Nazionale Galileo (TNG), the William Herschel Telescope (WHT), the Nordic Optical Telescope (NOT), the Joan Oró Telescope (TJO) of the Montsec Astronomical Observatory (OAdM), the STELLA telescope, and on observations collected at Asiago observatory. The LBT is an international collaboration among institutions in the United States, Italy, and Germany. The TNG, WHT, and NOT are operated on the island of La Palma  at the Spanish Observatorio del Roque de los Muchachos of the Instituto de Astrofisica de Canarias (IAC). The TNG is operated by the Fundación Galileo Galilei of the Istituto Nazionale di Astrofisica (INAF). The WHT is operated by the Isaac Newton Group and is run by the Royal Greenwich Observatory at the Spanish Roque de los Muchachos Observatory in La Palma. The NOT is operated jointly by Denmark, Finland, Iceland, Norway, and Sweden. The TJO of OAdM is owned by the Catalan Government and operated by the 
Institute for Space Studies of Catalonia (IEEC). STELLA is owned by the Leibniz Institute for Astrophysics Potsdam (AIP), which is jointly operated by AIP and IAC. We are grateful for DDT observing time at the TNG and for the acceptance of a Fast-Track program at the NOT. We wish to acknowledge all technical support at these observing facilities. We thank Jonathan Fortney for providing the cloud-free solar-composition atmospheric model for HAT-P-12b and the anonymous referee for many thoughtful comments, which significantly improved the manuscript.
\end{acknowledgements}

\bibliographystyle{aa}
\bibliography{mybib}

%
%
%

\end{document}